# A Primer on Computational Simulation in

# Congenital Heart Disease for the Clinician

Irene E. Vignon-Clementel[1*], Alison L. Marsden[2], Jeffrey A. Feinstein[3,4]

[1] INRIA Paris-Rocquencourt, Rocquencourt, France; irene.vignon-clementel@inria.fr

[2] Mechanical and Aerospace Engineering Department, University of California, San Diego, La Jolla CA 92093, USA; amarsden@ucsd.edu

[3] Pediatrics and [4] Bioengineering Departments, Stanford University, Stanford, CA 94305, USA; jeff.feinstein@stanford.edu

_______________________
*Corresponding author. Email irene.vignon-clementel@inria.fr

## Abstract

Interest in the application of engineering methods to problems in congenital heart disease has gained increased popularity over the past decade. The use of computational simulation to examine common clinical problems including single ventricle physiology and the associated surgical approaches, the effects of pacemaker implantation on vascular occlusion, or delineation of the biomechanical effects of implanted medical devices is now routinely appearing in clinical journals within all pediatric cardiovascular subspecialties. In practice, such collaboration can only work if both communities understand each other's methods and their limitations. This paper is intended to facilitate this communication by presenting in the context of congenital heart disease (CHD) the main steps involved in performing computational simulation - from the selection of an appropriate clinical question/problem to understanding the computational results, and all of the "black boxes" in between.

We examine the current state of the art and areas in need of continued development. For example, medical image-based model-building software has been developed based on numerous different methods. However, none of them can be used to construct a model with a simple "click of a button." The creation of a faithful, representative anatomic model, especially in pediatric subjects, often requires skilled manual intervention. In addition, information from a second imaging modality is often required to facilitate this process. We describe the technical aspects of model building, provide a definition of some of the most commonly used terms and techniques (e.g. meshes, mesh convergence, Navier-Stokes equations, and boundary conditions), and the assumptions used in running the simulations. Particular attention is paid to the assignment of boundary conditions as this point is of critical importance in the current areas of research within the realm of congenital heart disease. Finally, examples are provided demonstrating how computer simulations can provide an opportunity to "acquire" data currently unobtainable by other modalities, with essentially no risk to patients.

To illustrate these points, novel simulation examples of virtual Fontan conversion (from preoperative data to predicted postoperative state) and outcomes of different surgical designs are presented. The need for validation of the currently employed techniques and predicted results are required and the methods remain in their infancy. While the daily application of these technologies to patient specific clinical scenarios likely remains years away, the ever increasing interest in this area among both clinicians and engineers makes its eventual use far more likely than ever before and, some could argue, only a matter of [computing] time.

**Keywords:** hemodynamics; computer modeling; boundary conditions; clinical data; congenital heart disease

**Introduction**

> *Work in physiological fluid dynamics needs very close and intimate collaboration between specialists in physiological science and specialists in the dynamics of fluids. The necessary collaboration has to be preceded by a process of mutual education sufficiently prolonged to bring about on each side an adequate understanding of the other side's language and modes of expression, as well as recognition of which are the main areas where the other discipline has developed a particularly extensive and intricate body of knowledge and skills which can be called upon when required. After this, real communication between the different specialisms becomes possible, and can lead to effective research progress.*
> - *Sir James Lighthill[1]*

Over the past two decades, a similar paradigm has emerged, with collaboration among clinicians (adult and pediatric cardiologists, radiologist, surgeons, etc.) and scientists (mathematicians, engineers, physicists, computer scientists, etc.) toward the use of simulation based techniques and technologies in medicine. One obvious example is the application of computational fluid dynamics and computational simulation to the total cavopulmonary connection (Fontan operation). What likely started as a challenging and interesting set of homework assignments for engineering students has evolved, and recently emerged as a growing field of active, provocative and field-advancing research investigation. In practice, such collaboration can only work if both communities understand each other's methods and their limitations. This paper is thus intended to facilitate this communication, part of Sir Lighthill's "mutual education," by presenting, in the context of congenital heart disease (CHD), the main steps involved in defining an applicable clinical scenario/problem, building realistic computational models and understanding the virtues and pitfalls of such an approach. As final hemodynamic results are largely driven by the boundary conditions of the computer model (i.e. the parameters/restrictions put on the outlets of the specific model), special emphasis is put on how to incorporate patient-specific clinical data into the different types of boundary conditions most commonly used in simulation. The last section of the paper presents novel steps towards predictive modeling and surgical planning which this approach allows. It is important to realize these general ideas are also quite relevant for the study of structural dynamics as in the context of medical device design or blood and vessel wall dynamic interaction.

**General process from clinical input to computer simulation results**

In this section, we present a summary of the main steps to go from selecting a clinical area of study to obtaining and interpreting the computer simulation results.

*1.1    Selection of a clinical case*

The first step in the application of simulation-based techniques should come from interactions between a clinician and an engineer. The most efficient scenario is when the clinician comes with a clinical question that can be answered by the current methods available to the scientist (e.g. "Can you explain/confirm the reason for the pressure gradient in this patient?"). By attending clinical rounds and observing diagnostic testing techniques, and treatment methods to gain familiarity with the clinical decision making process, the scientist may also come up with a hemodynamics hypothesis to be tested via simulations.

In all cases, the selection of an "interesting problem" can only come from the joined will of the clinician and the scientist to bring together their areas of expertise. Together they will formulate questions that must be "simple enough" to be answered by computer modeling. To define a clinically relevant question, and select the appropriate scientific methods to answer it, is often an iterative process. Such a process, when successful, leads to novel methods and information in both communities.

Often, a hypothesis can first be tested with generic data whereby an idealized geometry is defined based on the clinician's knowledge of a "typical" anatomy for the selected medical condition, and physiological conditions (the functional information) come from literature data or typical clinical values provided by the clinician (representative references in CHD include [2-4]). Following this, hypotheses must be translated to the patient specific setting to evaluate robustness and clinical relevance[5-9] (see section 4 for such examples). Ultimately, a multi-patient longitudinal study will confirm or deny the hypothesis, and lead to follow up studies[10-13]. This last step although crucial, necessitates sufficient access to the relevant clinical data.

In this article, we focus on the patient-specific context, for which three-dimensional computer simulations are designed to answer the selected clinical question. Given the huge number of blood vessels in the cardiovascular system and the limited number of vessels that can be seen in conventional imaging data, a region of interest is defined for all problems. It typically includes a combination of the following (depending on the type of congenital heart disease): heart chambers, AV and/or semilunar valves, the great vessels, anastomoses (e.g. as seen in coarctation repairs), grafts (e.g. Fontan or modified Blalock-Taussig shunts), stents or other medical devices.

## 1.2   *From imaging data to geometric files readable by a numerical solver*

The most common structural imaging modalities clinically used in congenital heart disease are echocardiography, catheterization based angiography, magnetic resonance angiography (MRA), and computerized tomography (CT). To reconstruct accurate three-dimensional models that represent the anatomy of the patient, MRA and CT are usually the preferred imaging modalities, since the other two modalities do not provide the same level of three-dimensional information.

Three types of reconstruction exist. The first one is based on a two dimensional segmentation[14-16]. The segmentation[14-16]. The main steps, as shown in

Figure 1 are: 1) creation of vessel paths (which yields a stick-figure like structure of the patient anatomy), 2) segmentation of vessel lumens in the different MRA slices (a 2-dimensional representation of the vessel at a particular "slice" in the data), 3) creation of a 3-dimensional representation of each blood vessel wall by lofting these contours, and 4) assembly of these vessels into a three-dimensional surface model (the "geometrical model"). These different steps can be done with variants depending on the software, with more or less automation as dictated by the resolution of the data (e.g. [17]).

A less time consuming and more informative alternative is direct, three-dimensional segmentation[18, 19], as is used in commercial and open source software packages such as Mimics and ITK Snap.  These methods use intensity thresholding where the imaging area to be used in the model is automatically detected based on a defined voxel intensity or intensity gradient. This method is less robust for noisy data, especially for the younger/smaller pediatric subjects where the clinical MRA resolution may not be as high (for example, patients status post Norwood or Glenn operations where the primary vessels of interest are the pulmonary arteries which are not easily distinguishable from their venous partners).

A third method is based on predefined shapes (tubes, junctions, etc.) that are scaled, transformed and assembled to match the anatomy of a specific patient[20-22]. However, this is not very common in congenital heart disease, because of the anatomic variability and resolution issues. If possible, the vessel (or device) wall thickness is extracted with similar methods. This method is primarily applied in structural studies, as opposed to hemodynamic studies, for which incorporating this information plays a less important role.

**Take home points:** Commercial as well as in-house model-building software has been developed based on numerous different methods. However, none of them can be used to construct a model with a simple "click of a button." The creation of a faithful, representative anatomic model, especially in pediatric subjects, often requires skilled manual intervention. In addition, information from a second imaging modality may facilitate this process (e.g. MRA based models are often "validated" by closely examining cine angiogram data to insure all appropriate vessel branches or structural details have been included, Figure 2).

### 1.3 *"Running simulations": from preprocessing to understandable results*

Once the anatomy has been translated into a three-dimensional geometrical model, it is discretized (broken down) into a number of blocks (elements or volumes) connected by points (nodes) that define a so-called "mesh" (see Figure 3) which allows for the application of mathematical methods to each element rather than to the model as a whole. The number of blocks or nodes depends on the level of refinement appropriate for the study. If the number of nodes is too small, the results will be significantly affected and likely inaccurate. If "too many" nodes are included, then high computational cost may make the problem unsolvable, depending on the available computational resources. High computational cost problems usually necessitate supercomputer-based parallel processing to obtain solutions in a reasonable amount of time for clinical applicability. Mesh convergence is a technique which balances simulation accuracy and computing resources (think Goldilocks and the three bears – not too big, not too small… just right). The literature will often include a statement such as, "the mesh was refined until "mesh convergence" was achieved" (for references on mesh generation methods for blood flow simulations, see the recent review from Taylor and Steinman[15]).

To study blood flow using computational methods, we use the Navier-Stokes equations to represent the three-dimensional motion of blood in a continuous domain ($\Omega$) over a period of time (time zero to $T$). Considering blood as a Newtonian fluid, we formulate the balance of mass and momentum as follows:

Given $\vec{f} : \Omega \times (0,T) \rightarrow \Re^3$, find $\vec{v}(\vec{x},t)$ and $p(\vec{x},t)$ $\forall \vec{x} \in \Omega$, $\forall t \in (0,T)$ such that:

$$\nabla \cdot \vec{v} = 0$$

$$\rho \frac{\partial \vec{v}}{\partial t} + \rho (\nabla \vec{v}) \vec{v} = -\nabla p + \mu \Delta \vec{v} + \vec{f} \tag{1}$$

The primary variables are the blood velocity $\vec{v} = (v_x, v_y, v_z)$ and the pressure $p$ that vary in space $x, y, z$ and time t. The blood density, $\rho$, and the dynamic blood viscosity, $\mu$, are assumed constant in space and time, and rarely measured in an individual patient. Commonly used values are 1.06 g/cm$^3$ and 0.04 Poise, respectively. Note that taking into account non-Newtonian effects is possible, provided that the above equations are changed appropriately.

The external force $\vec{f}$ is usually zero, but it can be included if gravity or magnetic effects are taken into account.

Significant advances have recently been made recently in the area of fluid-structure interaction (FSI), methods which allow the determination of the amount of vessel wall deformation as blood passes through it or as a result of the effects of muscle and soft tissue surrounding them. Prior to the development of these methods, all simulations used rigid wall assumptions, i.e. vessels were modeled as if they were stiff and did not deform with changes in pressure. FSI simulations require coupled fluid-solid equations to be solved. Recent work has demonstrated FSI methods for the Fontan surgery, ventricular mechanics, cerebral aneurysms, and other applications[23-35]. FSI can ideally provide a more accurate and physiologic description of the hemodynamics as well as vessel wall stresses and strains. However, this comes at the cost of requiring more customized numerical methods, higher computational cost, and determination of vessel wall properties, which are often unknown, especially on a patient-specific basis. Whether or not wall motion is included, the equations can only be solved if appropriate initial and boundary conditions are given.

Assuming these conditions are specified, the Navier-Stokes equations are discretized (i.e. broken down into "digestible" pieces) in space and time with a chosen numerical method (the most common families are based on "finite elements" and "finite volumes"), which each have advantages and disadvantages and have been an important subject of research in the last fifty years. These discretized equations are then solved on a computer at each entity of the mesh.

The physical time scale of flow phenomema studied in a typical simulation is on the order of seconds (ranging from a couple of seconds to study cardiac pulsatile effects to dozens of seconds to study respiratory effects). This time scale should be distinguished from the "wall clock" time it takes to run the simulation on a computer, which depends on the numerical method, the desired accuracy and the available computational resources. For example, to run a typical patient-specific Navier-Stokes simulation with a one-million-element mesh for three cardiac cycles (which is often the minimum necessary to wash out initial transient effects), takes roughly one day on a sixty-processor machine using the latest in parallel processing technology.

Iterative procedures may be required before converging on a final simulation result. For example, mesh adaptation can be performed by iterating between mesh generation and equation solving to achieve mesh convergence of the results and better refinement of smaller vessels or regions of complex flow (Figure 3)[36]. Iterative procedures may also be necessary to tune the simulation results to match available clinical data[37], as discussed in the next section.

After satisfactory flow and pressure solutions are obtained, they are "post-processed" and visualized at selected locations. Often, other quantities of interest are computed from the basic flow fields, including blood flow repartition to the different branches, pressure gradient across a stenosis, energy losses in the model, wall shear stress, and pressure wave propagation. More sophisticated methods are being developed to better understand flow structures[38, 39] or how long particles (e.g. platelets) stay in a recirculation zone, which could explain thrombus formation. Discussions between the clinician and the scientist are then key to choose such quantities and to interpret the results.

**Take home points:** Geometric models are broken down into meshes, which allow for the application of mathematical methods to each element rather than to the model as a whole. Mesh convergence (remember Goldilocks) enables one to obtain an accurate solution with a mesh that is sufficiently dense and not overly demanding of computing resources. The Navier-Stokes equations are used to represent the three-dimensional motion of blood, and new methods allow for the modeling of the interaction between blood and the surrounding vessel (fluid-structure interaction, FSI) which, while computationally more demanding, are often considerably more physiologically relevant. In addition to flow and pressure results, CFD results often include analysis of flow distribution, shear stress analysis, and system efficiency determinants.

**Boundary conditions: a sensitive link to clinical data**

The previous section presented the general framework to build a geometric model and allow for mathematical methods to be used in the simulation. We now pay special attention to the boundary conditions, the input parameters, since they form the link between clinical data and simulated hemodynamics, and play a major role in determining the simulation results.

### *1.4    What is a boundary condition and where should it be placed?*

From a mathematical point of view, the equations presented above necessitate that some information is provided some information is provided at the boundaries of the domain/model (pressure, velocity, etc.). This constitutes This constitutes the formal definition of a boundary condition. From a practical standpoint, it is impossible to impossible to include all the vessels in the cardiovascular system in a three-dimensional hemodynamics hemodynamics simulation.  First, there is a limit in resolution (vessels below a certain size are not well are not well characterized from image data), and second, it is simply not computationally feasible (for hardware, feasible (for hardware, software, and time reasons). The 3D geometrical model is thus restricted to a domain of restricted to a domain of interest, as shown in the Glenn example of

Figure 1. The question of where to cut this domain is not trivial: the further the boundary is from the area of interest, the less influence this artificial boundary will have on the local results, but the larger the computational domain will be. In general engineering terms, there is a trade-off between accuracy and cost.

### *1.5    Why are proper boundary conditions needed?*

Contrary to other fluid mechanics applications, boundary conditions in blood flow simulations drive a large part of the dynamics inside the system under study. We illustrate this point with an example of flow and pressure simulated in a Fontan (TCPC) model in the simplest setting (resting steady conditions). Velocity vectors describing blood flow speed and direction are prescribed at the inlets (the superior vena cava, SVC, and inferior vena cava, IVC), whereas at the outlets, two different types of boundary conditions are prescribed. The first one is zero pressure (corresponding to an open vessel) and the second one is the simplest relationship between pressure and flow (their ratio), namely a resistance boundary condition, in this case proportional to the area of each vessel.

Despite the fact that this is a resting, steady regime, the resulting flow distribution between the left and the right lungs varies significantly with the type of boundary condition, as shown in Figure 4. Patient-specific blood flow distribution cannot be taken into account when imposing the same pressure at all the outlets. Furthermore, this has a direct impact on the velocity field (see Figure 5), and thus on the wall shear stress, as well as on the level of pressure in the system (which is particularly important for fluid-solid interaction applications) and the energy losses. This example illustrated the impact of outlet boundary conditions, but several studies have also demonstrated that taking into account the (cardiac[40] or respiratory[41]) pulsatility of the inflow as opposed to prescribing a steady value also significantly affects the results. A more detailed discussion on the impact of boundary condition types on the results, including pressure wave propagation in pulsatile settings can be found in works of Vignon-Clementel et al and Ballosino et al.[42-44]

### 1.6    How to incorporate clinical data in boundary conditions?

Patient-specific blood flow imaging (phase-contrast magnetic resonance imaging - PC-MRI, ultrasound, etc.) and catheterization based pressure acquisition technologies have increased in sophistication over the past two decades. These advances have triggered the development of new boundary conditions and enabled the incorporation of more patient-specific clinical data in computer simulations. Several examples are presented below, specific to the context of congenital heart disease and single ventricle patients, but the reader should keep in mind that similar methods are employed for other congenital and acquired disease simulations.

#### 1.6.1    Inlet boundary conditions

Hemodynamic conditions are typically simulated in multi-branched geometries, where blood enters through one vessel (the aorta for coarctations, the SVC for Glenn patients, the main pulmonary artery for Tetralogy of Fallot, etc) and more rarely two vessels (such as the SVC and the IVC in Fontan repairs). Blood flow at these locations is generally measurable with PC-MRI or ultrasound. The velocity field directly measured by PC-MRI under physiological conditions, such as rest and exercise (see e.g. [45, 46]), usually cannot be directly applied as a boundary condition. This is due to spatial resolution issues exacerbated in pediatric subjects. The PC-MRI velocities are thus integrated to obtain the flow rate as a function of time, which is then mapped to a chosen velocity profile (flat, parabolic or corresponding to Womersley theory depending on the vessel).

If the MRI acquisition is real-time, the effect of respiration can be included if necessary, provided the acquisition is long enough to accurately capture this information. However clinical acquisitions are often cardiac-gated and the resulting signal does not contain the respiratory variation. To include the main effect of respiration, which is particularly important in the IVC of Fontan subjects (see Figure 6, as well as [47]), a respiratory component is added to the cardiac flow rate[5]. This component can be computed based on the patient respiratory rate, if recorded during the acquisition. This is particularly easy to obtain for the younger subjects under artificial ventilator.

To include the normal or pathophysiological aperiodicity of the flow rate, information from cardiac-gated MRI and routinely acquired ultrasound (sonogram such as in Figure 6) can be combined to take advantage of both modalities (see for example the Glenn case in [8]).

Note that time-varying pressure signals are usually available from catheterization, but they are not prescribed as inlet boundary conditions as the amount of blood flowing in the computer model is very sensitive to pressure gradients between inlet(s) and outlets. These are not currently acquired with enough precision to serve this purpose.

#### 1.6.2    Outlet boundary conditions

While it is typically possible to obtain time-varying inflow rates from direct clinical measurements, it is often much harder to obtain flow rate information at the outlets of a multi-branched model. Resolution issues increase in these smaller vessels, especially in pediatric subjects, and additional scans to acquire multiple outlet flows require unrealistically long scan times. Furthermore, as these different time-varying signals are not acquired simultaneously, their synchronization is not obvious. While one may be tempted to use a mean constant pressure value as an outflow boundary condition, this has been shown to lead to significantly altered flow and pressure distributions[8, 44].

Given the above issues, boundary conditions that prescribe a relationship between pressure and flow are the preferred method to represent the response of vascular beds downstream of each outlet. These boundary conditions include resistance, impedance, Windkessel (RCR) and other lumped parameter models. This type of boundary condition allows for the prediction of hemodynamic changes in physiological conditions (such as respiration or exercise) or in surgical design, that may alter pressure or flow as measured during the clinical exam. A more involved discussion on the influence of boundary condition types on the simulated hemodynamics can be found in Vignon-Clementel et al. 2006[44].

The degree of patient-specificity in boundary conditions largely depends on the available clinical data. In congenital heart disease, PC-MRI and catheterization are routinely performed clinically, but the standard of care varies among centers.  For example, prior to Fontan repair, a typical MRI study would involve acquiring the SVC flow but not necessarily the right and left pulmonary flows.  In a simulation study, this information is crucial for determining downstream pulmonary flow and resistances. As seen in section 3.2, a lack of patient specific information about pulmonary flow split and the application of generic pressure boundary conditions significantly alters the hemodynamics for the entire system (including energy losses). Studies have also shown that the pulmonary flow distribution varies greatly from patient to patient.  For example in a simulation study of 6 Fontan patients, it varied from 54 to 70% (% of flow to the right lung)[10]. In these cases, there is a clear demonstration that boundary conditions must be carefully tuned to match clinical data to obtain reliable simulation results.

Catheterization data is an important component of boundary condition tuning, and can provide valuable additional information beyond the PC-MRI data. While many centers do not routinely perform pre-Fontan catheterization studies, this information should be included when available. For example, the overall left and right pulmonary resistances should be computed to match patient-specific flow split but should also be coherent with the mean pressures measured in the main vessels.  If the catheterization report indicates 10 mmHg in the SVC and in both PAs, then a "coherent" boundary condition should lead to a pressure drop of less than 1mmHg between these different vessels. If discrepancies exist, then additional tuning may be required to reconcile differences in the data and clinical measurements, and this should take into account the relative confidence in different clinical measurements.

When detailed information about smaller branch vessels is not available, then approximations must be made to choose appropriate conditions for each. For example, relative resistances of the smaller branch vessels can be made proportional to their respective outlet areas[5, 6, 8, 10]. Approximations may also be made using Murray's law (which relates the radii of child branches to the radii of the parent branch) or morphometry data, when available.

While resistance boundary conditions are straightforward and easier to implement, more accurate physiologic behavior can be obtained using more sophisticated boundary conditions including impedance and circuit analog models. Impedance and Windkessel "RCR" models can require additional knowledge of the pressure pulse of the patient.[12]  They may incorporate morphometry data, as done in recent studies in which Windkessel models were tuned to find the best match to the impedance of a morphometric tree[6, 8, 10, 48]. Automatic iterative adaptation of the boundary conditions[49] and their sensitivity to patient-specific pressure and flow data are the subject of ongoing research.

### 1.6.3    Coupling of the model with the entire circulation

To predict the global adaptation of the circulatory system (heart rate changes, etc.) to modifications in the cardiovascular geometry (e.g. the removal of a coarctation or an increase in shunt size), a simplified model of the heart can be coupled to the inflow vessel in the three-dimensional domain (e.g the aorta). The advantage is that these reduced models (zero dimensional electrical analog where for example valves are represented by diodes) are easily adaptable to specific congenital heart disease topologies (including unusual shunts, etc.)[2, 50, 51]. Similarly, to take into account with more precision the response of the downstream vascular trees, extensive simplified "zero dimensional" models have been developed and coupled to the three-dimensional model[51, 52]. When this coupling is done at all the inlets and outlets, the three-dimensional model is thus linked to a simplified model of the entire circulation, which permits to study their interaction[2, 52-55]. In some cases however, it is rather the three-dimensional model that is used (e.g. stenosis or pressure losses) to compute a few input parameters of the zero-dimensional model (e.g. resistances)[56-58]. However the inverse problem of going from patient specific data to the specification of their many parameters is a current challenge.

### 1.6.4    Numerical challenges

Patient-specific hemodynamics simulations have called for a special attention to boundary conditions, and during their development numerical challenges have arisen. In this subsection, we discuss some of the main numerical issues and solutions encountered in patient-specific hemodynamics simulations (not restricted to CHD).

Resistance, Windkessel or more extensive reduced models are all zero-dimensional models, in that they do not contain spatial variables. This is in contrast to three-dimensional models where the time-varying values of pressure and flow are obtained at all locations in space, thus providing information on local hemodynamics. The "coupled multidomain method" provides a theoretical derivation of the coupling of different models[43, 44]. Other means of coupling 0D and 3D models have been devised, and the dimensional mismatch of such coupling, which leads to an ill-posed problem, has been investigated [29, 30, 59-62]. For example, when pressure information is sent from the 3D to the 0D model, pressure $p^{3D}$ averaged over the coupling boundary section is sent as an input to the 0D model. In turn, solving the equations of the 0D model will provide a flow rate $Q^{0D}$, which must be applied at the 3D boundary. This transfer of information requires information in the 3D model, such as the velocity profile, which is not provided by the 0D model. There are several ways to numerically solve these different equations. It can be done all at once using a "monolithic" approach or by an iterative procedure. Variants on these methods, along with their advantages and disadvantages, have been the subject of intense research in the last decade in the applied mathematics & computational mechanics communities.

More recently, with increasing mesh resolution, more extensive computing power and increasingly physiological simulations, (including respiratory effects, exercise, etc.), complex flow structures are better captured in simulations. Complex flow has also been observed in-vivo and in-vitro[63-65]. The flip side is that they have created new numerical challenges. Complex flow tends to produce numerical instabilities and divergence of solutions, rarely mentioned in publications, and which exact causes and remedies are subject to current research. Several solutions have been proposed so far. These include modeling enough branches such that flow is laminar and sufficiently unidirectional at boundaries[66], constraining the velocity profile at the coupling interfaces[67], addition of stabilization terms for areas of reversed flow[68]. Congenital heart disease tends to present complex anatomies that generate intricate flow fields, such as stenoses (e.g. coarctation, pulmonary stenosis), shunts (e.g. Norwood repair), artificial junctions (e.g. Glenn & Fontan repairs), etc. Thus, tackling these numerical challenges is particularly relevant for a better understanding and treatment of these diseases.

**Take home points:** Hemodynamics in the region of interest cannot be simulated in isolation from the rest of the cardiovascular system. The choice of boundary conditions must be carefully considered for each problem, and relevant clinical and physical information must be included at each inflow and outflow boundary.

**Steps towards predictive modeling**

Three-dimensional simulations can provide information which are difficult or impossible to obtain in the clinic, such as an entire velocity or pressure field, wall shear stress, energy losses, etc. Such information usually describes the current, resting state of the patient. An even more powerful use of computer simulations is to predict hemodynamic conditions in altered physiologic or surgical states. This capability necessitates careful modeling of this state change. The next subsections provide a few illustrative examples of the use of simulations for predictive modeling using patient specific data.

*1.6.5   Validation of the results with clinical data*

A crucial aspect of computer simulations is their validation. Verification of a simulation ensures that the equations are solved correctly. This is a natural part of the development of computational tools. By contrast, validation ensures that the numerical results are "faithful" to reality. The results come from a model, which by nature, is based on simplifying hypotheses. This model must represent physical reality with enough precision so that the answer to desired clinical question is robust and meaningful. Existing tools for validation include comparison of the results to in-vitro data[63, 64, 69-71] using PIV and dye visualization with stand-alone models or mock-circuits, to in-vivo measurements in animals[48] or in humans[65]. In the latter, for example, velocity results can be compared to MRI data at a slice that was not used to create the numerical results[72].

Predicted pressures can be compared with catheterization data (when not used to set up the model)[52]. Figure 7 shows another example, in which the pressure waveform computed in the SVC of a Glenn patient was qualitatively compared to the one measured in the cath-lab[8]. Cine angiography movies can provide qualitative validation of the velocity dynamics even if quantitative data is not available.

Validation of all these methods (CFD, in-vitro and in-vivo measurement methods) is much needed[73] and is thus an important current research front[15], which requires scientists and clinicians to define the desired accuracy of the model and to identify the data necessary for validation. While this may require the acquisition of extra-data, this additional "cost" is critical if we are to rely on simulation based results in the future. The cost-benefit analysis of validation studies, which may require additional risk to a few patients for the benefit of many, is an important ethical topic worthy of discussion. In the context of predictive simulations, validation on a subset of cases is important and may call for novel clinical protocols, such as the acquisition of pressure measurements under exercise condition in Fontan patients, or post-operative MRI acquisition and lung perfusion studies.

*1.6.6    An example of virtual Fontan conversion: from preoperative data to predicted postoperative state*

In this example, clinical data was acquired from PC-MRI, catheterization and cine angiography, prior to Fontan angiography, prior to Fontan repair of a 3 year old male Glenn patient. The pre-operative data served as inputs served as inputs for a three-dimensional simulation, with geometry and boundary conditions constructed as constructed as explained in the above sections. For the geometry, the 2D-based segmentation was used as was used as described in

Figure 1. At the inlet, aperiodic flow was prescribed based on MRI and ultrasound data, whereas at the outlet, RCR boundary conditions were designed - see [8] for more details. A virtual surgery was then performed: the Glenn geometry was modified to include the IVC anastomosis with an extracardiac graft, the size of which was taken from the Fontan surgical report. The Fontan simulation was then performed in attempt to predict the post-operative hemodynamics. Identical boundary conditions were utilized in both the Glenn and Fontan states, with the addition of the new IVC inlet for which velocities were constructed using the pre-operative MRI flow data. Figure 8 shows the two geometries, as well as the resulting transpulmonary pressure gradients. Simulated transpulmonary pressure gradient for the preoperative state matches well with the value reported in the preoperative report. This is to be expected, since its boundary conditions were constructed to match the clinical data. By contrast, the simulated postoperative value is a predicted postoperative value, and it also matches very well the operative report value. The transpulmonary gradient almost doubled between preoperatively and postoperatively, which is consistent with the flow change.

The prediction is based on a model of the virtual surgery that only required a simple geometrical change, with the assumption that the pre-operative boundary conditions did not change postoperatively. Such a successful prediction needs to be validated in multiple cases to assess its robustness, since this simple virtual surgery model seems unlikely to hold in general. However, this result provides a promising example of the power of simulations to aid in surgical planning and outcome predictions.

*1.6.7    Predicting the outcome of different surgical designs*

Once a simulation model has been established that reliably represents the current state, modification of the state can be explored using virtual surgery. This example investigates the evaluation of competing designs for the Fontan surgery using virtual surgery tools. A novel Y-graft design has recently been proposed to replace tube-shaped grafts in the extra-cardiac Fontan procedure.[4, 6] Starting from a patient-specific model of the Glenn surgery, competing designs for the Y-graft, offset, and traditional t-junction were constructed. Simulations were run at rest and exercise to compare performance in multiple categories. Figure 9 shows the different models that were tested in simulation, together with performance results for hepatic flow distribution and power loss. Three Y-graft designs were tested, with input from surgical colleagues. Results demonstrate that the Y-graft is a promising surgical method, but that not all Y-graft designs have adequate performance. These results underscore the need for patient-specific surgical planning that move away from current "one-size-fits-all" treatment paradigms.

To underscore the importance of boundary conditions in simulation predictions, Figure 10 compares the predicted hepatic flow distribution and power loss for two different surgical designs using steady flow vs. pulsatile flow at the two inflow faces. While it is not surprising that power loss is higher for unsteady flow cases[5] results also show differences in hepatic flow that are as high as 30%. These results show that care must be taken to properly define both inflow and outflow boundary conditions, and that simulation results should not be taken at "face value" without questioning the validity and uncertainty of boundary condition parameters.

**Take home points:** Computer simulations provide the opportunity to "acquire" data currently unobtainable by other modalities with essentially no risk to patients. Validation of current techniques and results is required and still in its infancy. The promise of providing patient-specific simulations of planned interventions and incorporate these results into clinical decision making is exciting but remains years off.

### Conclusions

Progress in computational methods, the application of these methods to problems in congenital heart disease, validation of simulation results and predictive modeling will only be made through strong interactions and cooperation between engineers and clinicians. First, engineers should incorporate the clinicians' expertise on data and their interpretation (resolution, uncertainties, etc). They can then design boundary conditions that lead to more robust conclusions. As discussed in this article, boundary conditions constitute a crucial aspect of performing clinically relevant computer simulations. They also lead to modeling, numerical, and validation challenges, for which specific methods are being developed, addressing the issues of parameter estimation, coupling with the rest of the circulation and complex flow handling. This allows the field to move towards predictive simulations with clinical value. In addition, engineers can then understand the surgical constraints to design meaningful virtual surgical geometries. Eventually, this feeds back to the clinics by providing innovative tools or new ways of thinking for congenital heart disease diagnosis and treatment planning.

### Acknowledgments

Dr. Vignon-Clementel was supported by a Leducq Foundation Network of Excellence grant, a Stanford/France grant and an INRIA collaboratory grant. Dr. Marsden was supported by a Burroughs Wellcome Fund Career Award at the Scientific Interface, a Leducq Foundation Network of Excellence grant, and an American Heart Association Beginning Grant in Aid. Dr. Feinstein was supported by the Vera Moulton Wall Center at Stanford University. The authors gratefully acknowledge the input of clinical colleagues Dr. F.P. Chan (Radiology, Stanford Hospital), Dr. V. Mohan Reddy (Cardiothoracic Surgery, Stanford University), Dr. John Lamberti (Rady Children's Hospital San Diego), and Dr. Tain-Yen Hsia (Great Ormond Street Hospital, London) and engineering colleague Dr. C.A. Taylor (Departments of Bioengineering, Surgery, Mechanical Engineering and Radiology, Stanford University). Software was generously provided by the SimVascular open source project at Stanford University (simtk.org)[74].

## References

1.  Lighthill J. *Mathematical biofluiddynamics*. Philadelphia; 1975.

2.  Bove E, Migliavacca F, de Leval M, Balossino R, Pennati G, Lloyd T, Khambadkone S, Hsia T, Dubini G. Use of mathematic modeling to compare and predict hemodynamic effects of the modified blalock-taussig and right ventricle-pulmonary artery shunts for hypoplastic left heart syndrome. *Journal of Thoracic and Cardiovascular Surgery*. 2008:312-320

3.  Soerensen DD, Pekkan K, de Zelicourt D, Sharma S, Kanter K, Fogel M, Yoganathan AP. Introduction of a new optimized total cavopulmonary connection. *Annals of Thoracic Surgery*. 2007;83:2182-2190

4.  Yang WG, Feinstein JA, Marsden AL. Constrained optimization of an idealized y-shaped baffle for the fontan surgery at rest and exercise. *Computer Methods in Applied Mechanics and Engineering*. 2010;199:2135-2149

5.  Marsden AL, Vignon-Clementel IE, Chan FP, Feinstein JA, Taylor CA. Effects of exercise and respiration on hemodynamic efficiency in cfd simulations of the total cavopulmonary connection. *Annals of Biomedical Engineering*. 2007;35:250-263

6.  Marsden AL, Bernstein AJ, Reddy VM, Shadden SC, Spilker RL, Chan FP, Taylor CA, Feinstein JA. Evaluation of a novel y-shaped extracardiac fontan baffle using computational fluid dynamics. *Journal of Thoracic and Cardiovascular Surgery*. 2009;137:394-U187

7.  Sundareswaran K, de Zélicourt D, Sharma S, Kanter K, Spray T, Rossignac J, Sotiropoulos F, Fogel M, Yoganathan A. Correction of pulmonary arteriovenous malformation using image-based surgical planning. *JACC Cardiovasc Imaging*. 2009;2:1024-1030

8.  Vignon-Clementel IE, Figueroa CA, Jansen KE, Taylor CA. Outflow boundary conditions for 3d simulations of non-periodic blood flow and pressure fields in deformable arteries. *Computer Methods in Biomechanics and Biomedical Engineering*. 2010;13(5) :625-640

9.  Tang D, Yang C, Geva T, Del Nido PJ. Patient-specific mri-based 3d fsi rv/lv/patch models for pulmonary valve replacement surgery and patch optimization. *J Biomech Eng*. 2008;130:041010

10. Marsden A, Reddy V, Shadden S, Chan F, Taylor C, Feinstein J. A new multiparameter approach to computational simulation for fontan assessment and redesign. *Congenit Heart Dis*. 2010;5:104-117

11. Whitehead KK, Pekkan K, Kitajima HD, Paridon SM, Yoganathan AP, Fogel MA. Nonlinear power loss during exercise in single-ventricle patients after the fontan - insights from computational fluid dynamics. *Circulation*. 2007;116:I165-I171

12. LaDisa Jr. JF, Taylor CA, Feinstein JA. Aortic coarctation: Recent developments in experimental and computational methods to assess treatments for this simple condition. *Progress in Pediatric Cardiology*. 2010

13.    Capelli C, Taylor AM, Migliavacca F, Bonhoeffer P, Schievano S. Patient-specific reconstructed anatomies and computer simulations are fundamental for selecting medical device treatment: Application to a new percutaneous pulmonary valve. *Philosophical Transactions of the Royal Society a-Mathematical Physical and Engineering Sciences*. 2010;368:3027-3038

14.    Steinman DA. Image-based computational fluid dynamics modeling in realistic arterial geometries. *Annals of Biomedical Engineering*. 2002;30:483-497

15.    Taylor CA, Steinman DA. Image-based modeling of blood flow and vessel wall dynamics: Applications, methods and future directions. *Annals of Biomedical Engineering*. 2010;38:1188-1203

16.    Wang KC, Dutton RW, Taylor CA. Improving geometric model construction for blood flow modeling - geometric image segmentation and image-based model construction for computational hemodynamics. *Ieee Engineering in Medicine and Biology Magazine*. 1999;18:33-39

17.    Frakes DH, Smith MJT, Parks J, Sharma S, Fogel M, Yoganathan AP. New techniques for the reconstruction of complex vascular anatomies from mri images. *Journal of Cardiovascular Magnetic Resonance*. 2005;7:425-432

18.    Antiga L, Piccinelli M, Botti L, Ene-Iordache B, Remuzzi A, Steinman D. An image-based modeling framework for patient-specific computational hemodynamics. *Medical & Biological Engineering & Computing*. 2008;46:1097-1112

19.    Bekkers EJ, Taylor CA. Multiscale vascular surface model generation from medical imaging data using hierarchical features. *Ieee Transactions on Medical Imaging*. 2008;27:331-341

20.    Sermesant M, Forest C, Pennec X, Delingette H, Ayache N. Deformable biomechanical models: Application to 4d cardiac image analysis. *Med Image Anal*. 2003;7:475-488

21.    Zhang YJ, BazilevS Y, GoswaMi S, Bajaj CL, Hughes TJR. Patient-specific vascular nurbs modeling for isogeometric analysis of blood flow. *Computer Methods in Applied Mechanics and Engineering*. 2007;196:2943-2959

22.    Ecabert O, Peters J, Schramm H, Lorenz C, von Berg J, Walker MJ, Vembar M, Olszewski ME, Subramanyan K, Lavi G, Weese J. Automatic model-based segmentation of the heart in ct images. *Ieee Transactions on Medical Imaging*. 2008;27:1189-1201

23.    Figueroa CA, Vignon-Clementel IE, Jansen KE, Hughes TJR, Taylor CA. A coupled momentum method for modeling blood flow in three-dimensional deformable arteries. *Computer Methods in Applied Mechanics and Engineering*. 2006;195:5685-5706

24.    Bazilevs Y, Calo VM, Zhang Y, Hughes TJR. Isogeometric fluid-structure interaction analysis with applications to arterial blood flow. *Computational Mechanics*. 2006;38:310-322

25.    Bazilevs Y, Hsu MC, Benson DJ, Sankaran S, Marsden AL. Computational fluid-structure interaction: Methods and application to a total cavopulmonary connection. *Computational Mechanics*. 2009;45:77-89

26.    Causin P, Gerbeau JF, Nobile F. Added-mass effect in the design of partitioned algorithms for fluid-structure problems. *Computer Methods in Applied Mechanics and Engineering*. 2005;194:4506-4527

27.    Gerbeau JF, Vidrascu M, Frey P. Fluid-structure interaction in blood flows on geometries based on medical imaging. 2005:155-165

28.    Badia S, Quaini A, Quarteroni A. Splitting methods based on algebraic factorization for fluid-structure interaction. *Siam Journal on Scientific Computing*. 2007;30:1778-1805

29.    Formaggia L, Gerbeau JF, Nobile F, Quarteroni A. On the coupling of 3d and 1d navier-stokes equations for flow problems in compliant vessels. *Computer Methods in Applied Mechanics and Engineering*. 2001;191:561-582

30.    Urquiza SA, Blanco PJ, Venere MJ, Feijoo RA. Multidimensional modelling for the carotid artery blood flow. *Computer Methods in Applied Mechanics and Engineering*. 2006;195:4002-4017

31.    Fernandez MA, Gerbeau JF, Martin V. Numerical simulation of blood flows through a porous interface. *Esaim-Mathematical Modelling and Numerical Analysis-Modelisation Mathematique Et Analyse Numerique*. 2008;42:961-990

32.    Torii R, Oshima M, Kobayashi T, Takagi K, Tezduyar TE. Computer modeling of cardiovascular fluid-structure interactions with the deforming-spatial-domain/stabilized space-time formulation. *Computer Methods in Applied Mechanics and Engineering*. 2006;195:1885-1895

33.    Chapelle D, Gerbeau J-F, Sainte-Marie J, Vignon-Clementel I. A poroelastic model valid in large strains with applications to perfusion in cardiac modeling. *Computational Mechanics*. 2010;46:91-101

34.    Watanabe H, Sugiura S, Kafuku H, Hisada T. Multiphysics simulation of left ventricular filling dynamics using fluid-structure interaction finite element method. *Biophys J*. 2004;87:2074-2085

35.    Nordsletten D, Niederer S, Nash M, Hunter P, Smith N. Coupling multi-physics models to cardiac mechanics. *Prog Biophys Mol Biol*. 2009

36.    Sahni O, Muller J, Jansen KE, Shephard MS, Taylor CA. Efficient anisotropic adaptive discretization of the cardiovascular system. *Computer Methods in Applied Mechanics and Engineering*. 2006;195:5634-5655

37.    Spilker R, Taylor C. Tuning multidomain hemodynamic simulations to match physiological measurements. *Ann Biomed Eng*. 2010;38:2635-2648

38.    Shadden S, Taylor C. Characterization of coherent structures in the cardiovascular system. *Annals of Biomedical Engineering*. 2008;36:1152-1162

39. Shadden SC, Astorino M, Gerbeau J-F. Computational analysis of an aortic valve jet with lagrangian coherent structures. *Chaos: An Interdisciplinary Journal of Nonlinear Science*. 2010;20:017512

40. DeGroff C, Shandas R. Designing the optimal total cavopulmonary connection: Pulsatile versus steady flow experiments. *Med Sci Monit*. 2002;8:MT41-45

41. Marsden AL, Vignon-Clementel IE, Chan FP, Feinstein JA, Taylor CA. Effects of exercise and respiration on the hemodynamic efficiency in cfd simulations of the total cavopulmonary connection. *Annals of Biomedical Engineering*. 2007;35:250-263

42. Balossino R, Pennati G, Migliavacca F, Formaggia L, Veneziani A, Tuveri M, Dubini G. Computational models to predict stenosis growth in carotid arteries: Which is the role of boundary conditions? *Computer Methods in Biomechanics and Biomedical Engineering*. 2009;12:113-123

43. Vignon IE, Taylor CA. Outflow boundary conditions for one-dimensional finite element modeling of blood flow and pressure waves in arteries. *Wave Motion*. 2004;39:361-374

44. Vignon-Clementel IE, Figueroa CA, Jansen KE, Taylor CA. Outflow boundary conditions for three-dimensional finite element modeling of blood flow and pressure in arteries. *Computer Methods in Applied Mechanics and Engineering*. 2006;195:3776-3796

45. Cheng C, Herfkens R, Lightner A, Taylor C, Feinstein J. Blood flow conditions in the proximal pulmonary arteries and vena cavae: Healthy children during upright cycling exercise. *Am J Physiol Heart Circ Physiol*. 2004;287:H921-926

46. Cheng C, Herfkens R, Taylor C, Feinstein J. Proximal pulmonary artery blood flow characteristics in healthy subjects measured in an upright posture using mri: The effects of exercise and age. *J Magn Reson Imaging*. 2005;21:752-758

47. Hjortdal V, Emmertsen K, Stenbøg E, Fründ T, Schmidt M, Kromann O, Sørensen K, Pedersen E. Effects of exercise and respiration on blood flow in total cavopulmonary connection: A real-time magnetic resonance flow study. *Circulation*. 2003;108:1227-1231

48. Spilker RL, Feinstein JA, Parker DW, Reddy VM, Taylor CA. Morphometry-based impedance boundary conditions for patient-specific modeling of blood flow in pulmonary arteries. *Annals of Biomedical Engineering*. 2007;35:546-559

49. Spilker RL. Computational analysis of blood flow in arteries incorporating reduced-order models of the downstream vasculature. 2009;PhD

50. Migliavacca F, Dubini G, Pennati G, Pietrabissa R, Fumero R, Hsia T, de Leval M. Computational model of the fluid dynamics in systemic-to-pulmonary shunts. *J Biomech*. 2000;33:549-557

51. Migliavacca F, Dubini G, de Leval M. Computational fluid dynamics in paediatric cardiac surgery. *Images Paediatr Cardiol*. 2000;2:11-25

52. Migliavacca F, Balossino R, Pennati G, Dubini G, Hsia T, de Leval M, Bove E. Multiscale modelling in biofluidynamics: Application to reconstructive paediatric cardiac surgery. *J Biomech*. 2006;39:1010-1020

53. Lagana K, Balossino R, Migliavacca F, Pennati G, Bove EL, de Leval MR, Dubini G. Multiscale modeling of the cardiovascular system: Application to the study of pulmonary and coronary perfusions in the univentricular circulation. *Journal of Biomechanics*. 2005;38:1129-1141

54. Hsia T, Migliavacca F, Pennati G, Balossino R, Dubini G, de Leval M, Bradley S, Bove E. Management of a stenotic right ventricle-pulmonary artery shunt early after the norwood procedure. *Ann Thorac Surg*. 2009;88:830-837; discussion 837-838

55. Pennati G, Migliavacca F, Dubini G, Bove EL. Modeling of systemic-to-pulmonary shunts in newborns with a univentricular circulation: State of the art and future directions. *Progress in Pediatric Cardiology*. This special issue

56. Pennati G, Migliavacca F, Dubini G, Pietrabissa R, de Leval M. A mathematical model of circulation in the presence of the bidirectional cavopulmonary anastomosis in children with a univentricular heart. *Medical Engineering & Physics*. 1997;19:223-234

57. Sundareswaran K, Pekkan K, Dasi L, Whitehead K, Sharma S, Kanter K, Fogel M, Yoganathan A. The total cavopulmonary connection resistance: A significant impact on single ventricle hemodynamics at rest and exercise. *Am J Physiol Heart Circ Physiol*. 2008;295:H2427-2435

58. Pekkan K, Frakes D, De Zelicourt D, Lucas C, Parks W, Yoganathan A. Coupling pediatric ventricle assist devices to the fontan circulation: Simulations with a lumped-parameter model. *ASAIO J*. 2005;51:618-628

59. Quarteroni A, Veneziani A. Analysis of a geometrical multiscale model based on the coupling of odes and pdes for blood flow simulations. *Multiscale Modeling & Simulation*. 2003;1:173-195

60. Formaggia L, Gerbeau J, Nobile F, Quarteroni A. Numerical treatment of defective boundary conditions for the navier-stokes equations. *SIAM Journal on Numerical Analysis*. 2002;40:376-401

61. Formaggia L, Gerbeau JF, Nobile F, Quarteroni A. On the coupling of 3d and 1d navier-stokes equations for flow problems in compliant vessels. *Computer Methods in Applied Mechanics and Engineering*. 2001;191:561-582

62. Blanco PJ, Feijoo RA, Urquiza SA. A unified variational approach for coupling 3d-1d models and its blood flow applications. *Computer Methods in Applied Mechanics and Engineering*. 2007;196:4391

63. de Zelicourt DA, Pekkan K, Wills L, Kanter K, Forbess J, Sharma S, Fogel M, Yoganathan AP. In vitro flow analysis of a patient-specific intraatrial total cavopulmonary connection. *Annals of Thoracic Surgery*. 2005;79:2094-2102

64.     Ryu K, Healy TM, Ensley AE, Sharma S, Lucas C, Yoganathan AP. Importance of accurate geometry in the study of the total cavopulmonary connection: Computational simulations and in vitro experiments. *Annals of Biomedical Engineering*. 2001;29:844-853

65.     Sharma S, Ensley AE, Hopkins K, Chatzimavroudis GP, Healy TM, Tam VKH, Kanter KR, Yoganathan AP. In vivo flow dynamics of the total cavopulmonary connection from three-dimensional multislice magnetic resonance imaging. *Annals of Thoracic Surgery*. 2001;71:889-898

66.     Vignon-Clementel IE. A coupled multidomain method for computational modeling of blood flow. *Mechanical Engineering*. 2006;Ph.D.

67.     Kim HJ, Figueroa CA, Hughes TJR, Jansen KE, Taylor CA. Augmented lagrangian method for constraining the shape of velocity profiles at outlet boundaries for three-dimensional finite element simulations of blood flow. *Computer Methods in Applied Mechanics and Engineering*. 2009;198:3551-3566

68.     Hsu MC, Bazilevs Y, Calo VM, Tezduyar TE, Hughes TJR. Improving stability of stabilized and multiscale formulations in flow simulations at small time steps. *Computer Methods in Applied Mechanics and Engineering*. 2010;199:828-840

69.     Dasi L, Krishnankuttyrema R, Kitajima H, Pekkan K, Sundareswaran K, Fogel M, Sharma S, Whitehead K, Kanter K, Yoganathan A. Fontan hemodynamics: Importance of pulmonary artery diameter. *J Thorac Cardiovasc Surg*. 2009;137:560-564

70.     Pekkan K, Dasi L, de Zélicourt D, Sundareswaran K, Fogel M, Kanter K, Yoganathan A. Hemodynamic performance of stage-2 univentricular reconstruction: Glenn vs. Hemi-fontan templates. *Ann Biomed Eng*. 2009;37:50-63

71.     Małota Z, Nawrat Z, Kostka P. Computer and physical modeling of blood circulation pump support for a new field of application in palliative surgery. *Int J Artif Organs*. 2007;30:1068-1074

72.     Migliavacca F, Kilner PJ, Pennati G, Dubini G, Pietrabissa R, Fumero R, de Leval MR. Computational fluid dynamic and magnetic resonance analyses of flow distribution between the lungs after total cavopulmonary connection. *IEEE Transactions on Biomedical Engineering*. 1999;46:393-399.

73.     DeGroff C. Modeling the fontan circulation: Where we are and where we need to go. *Pediatr Cardiol*. 2008;29:3-12

74.     Schmidt JP, Delp SL, Sherman MA, Taylor CA, Pande VS, Altman RB. The simbios national center: Systems biology in motion. *Proceedings of the Ieee*. 2008;96:1266-1280

## Figures and Legends

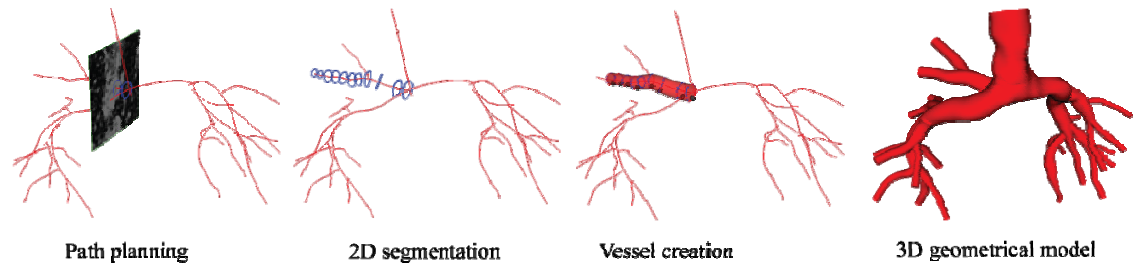

Figure 1: Main steps for the image-based anatomical construction, here with 2D segmentation of Glenn (bidirectional cavopulmonary anastomosis) MRA data.

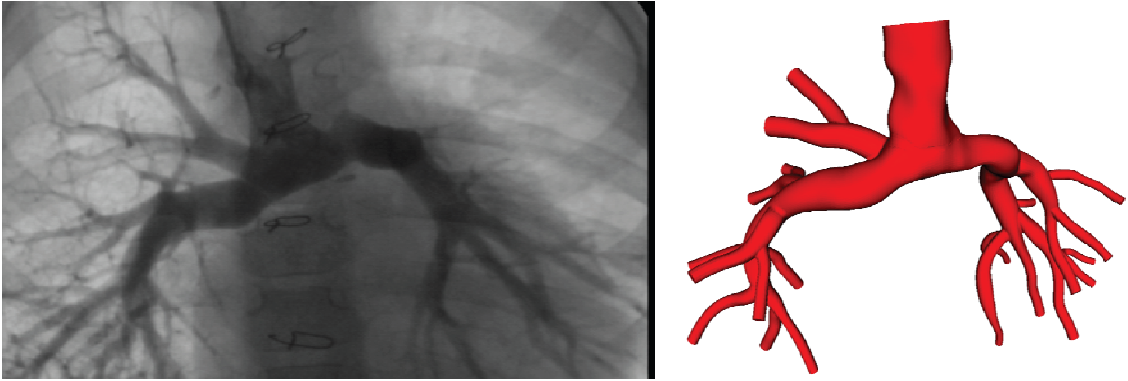

Figure 2: Example of cine-angiogram (left) that was used in combination with MRA data to confirm the three-dimensional geometrical model (right) contained all major pulmonary artery branches. Note that due to the nature of a cine-angiogram, the exact shape of each vessel may differ from the one seen on MRA.

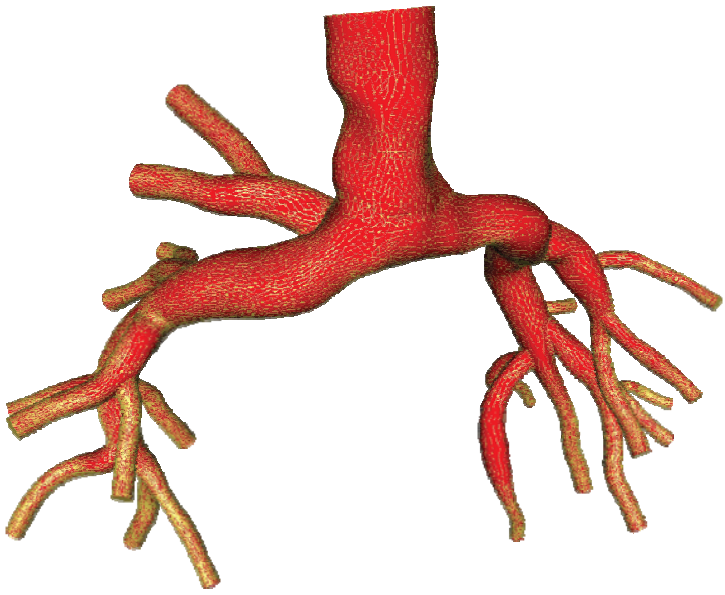

Figure 3: Anisotropic mesh (gold) on top of the anatomical model (red).

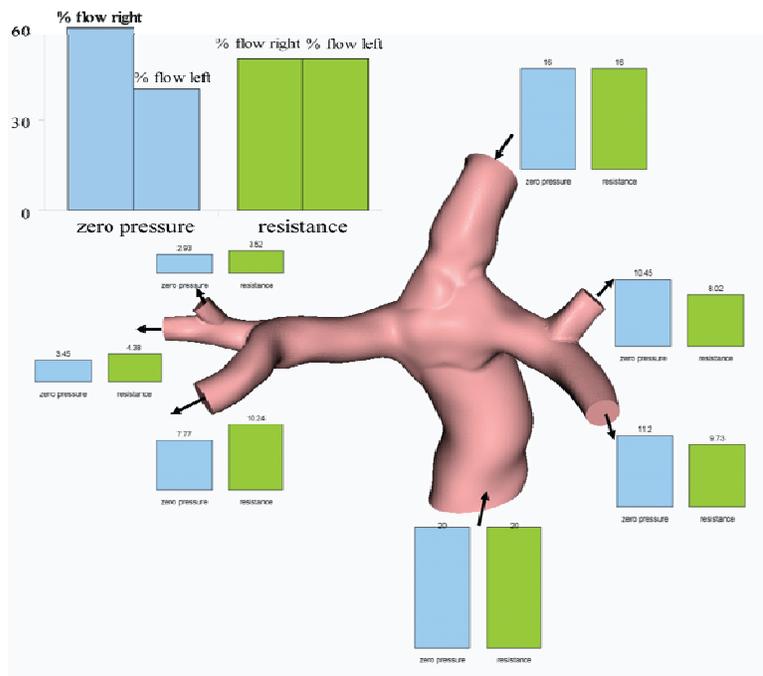

Figure 4: Top chart: overall lung flow distribution in a Fontan model depending on the various outlet boundary conditions. Around the 3D model are the detailed inflow and outflow distributions for these two conditions. Arrows indicate the direction of flow.

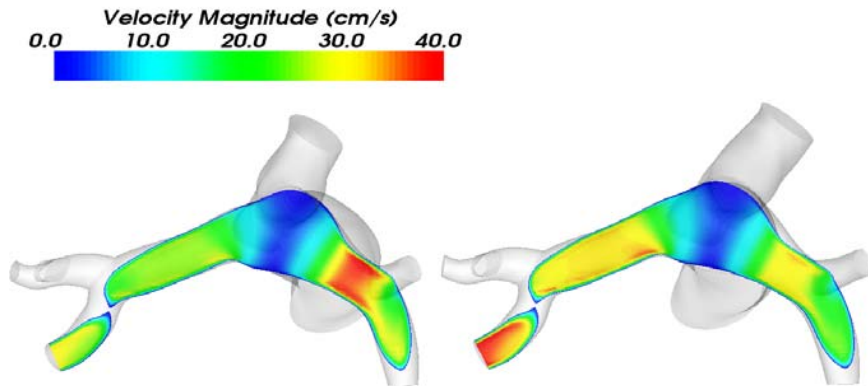

Figure 5: Velocity magnitude in the case of (left) zero pressure and (right) resistance outlet boundary conditions.

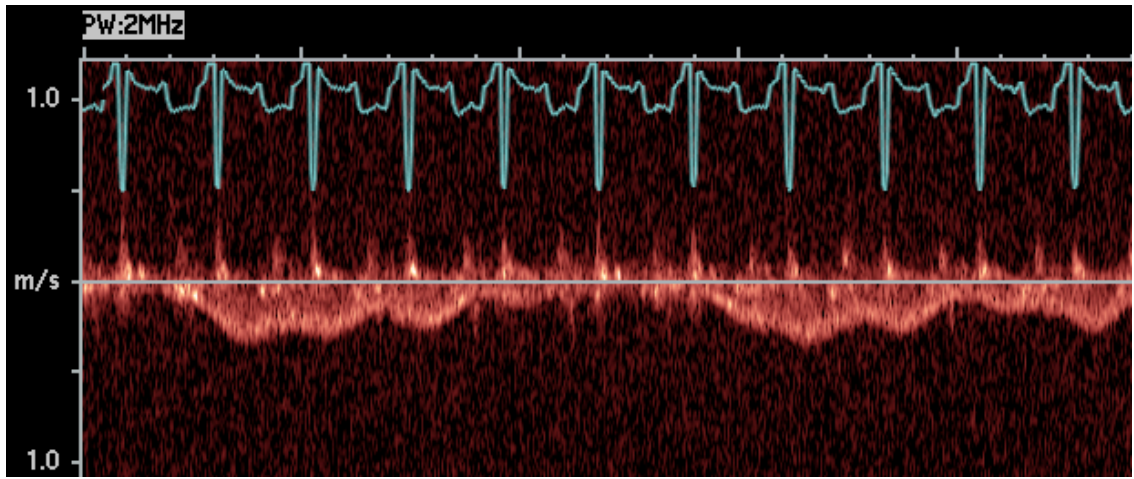

Figure 6: Doppler ultrasound demonstrating the time variability (including asynchronous cardiac and respiration effects) in the inferior vena cava of a Fontan patient shortly after repair.

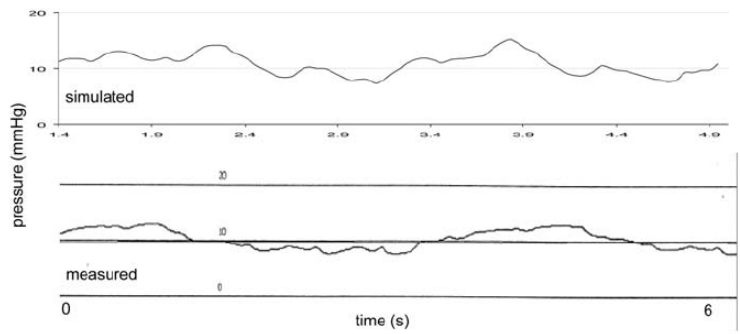

Figure 7: One example of simulation validation: comparison of simulated (top) and catheterization measured (bottom) pressures in the SVC of a Glenn patient. Reproduced from Vignon-Clementel et al.[8], with permission.

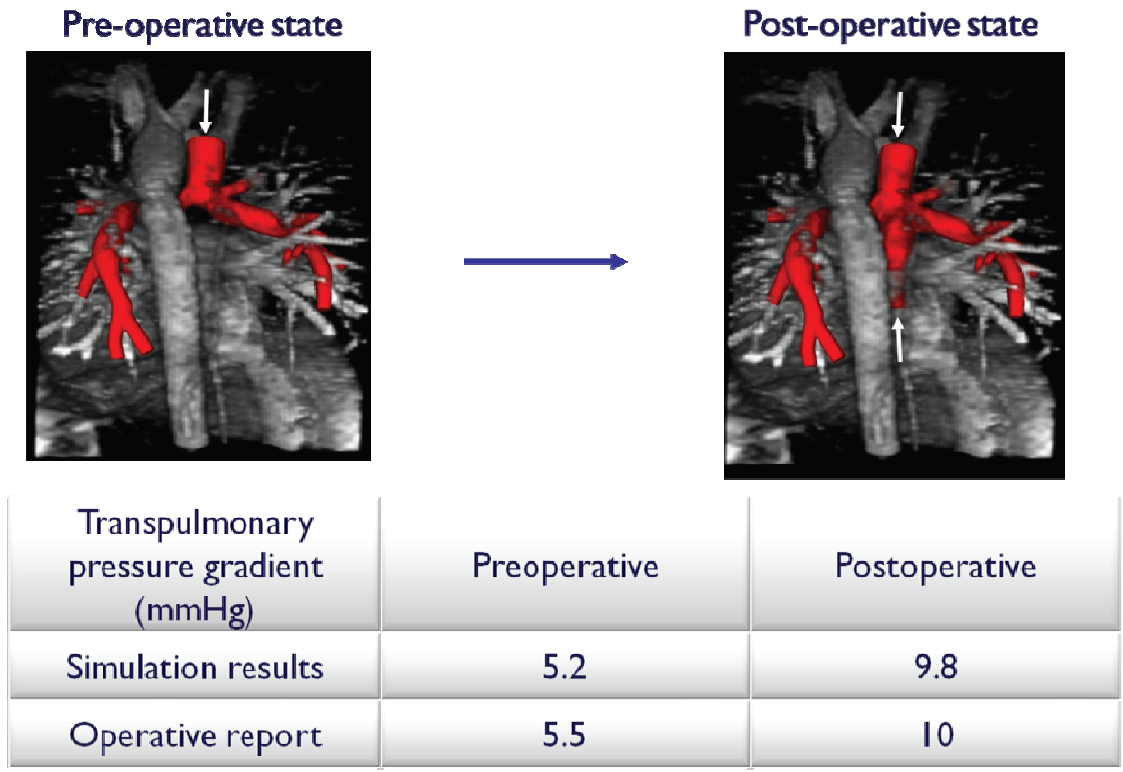

| Transpulmonary pressure gradient (mmHg) | Preoperative | Postoperative |
|---|---|---|
| Simulation results | 5.2 | 9.8 |
| Operative report | 5.5 | 10 |

Figure 8: Preoperative state and predicted postoperative state after virtual Fontan surgery. Top: Glenn geometry (left) and virtual Fontan geometry (right). White arrows indicate the inflow of each configuration. Bottom: simulated transpulmonary pressure gradient for the preoperative state (which boundary conditions were constructed to match the clinical data) and predicted postoperative value (comparison is with operative report pressure values).

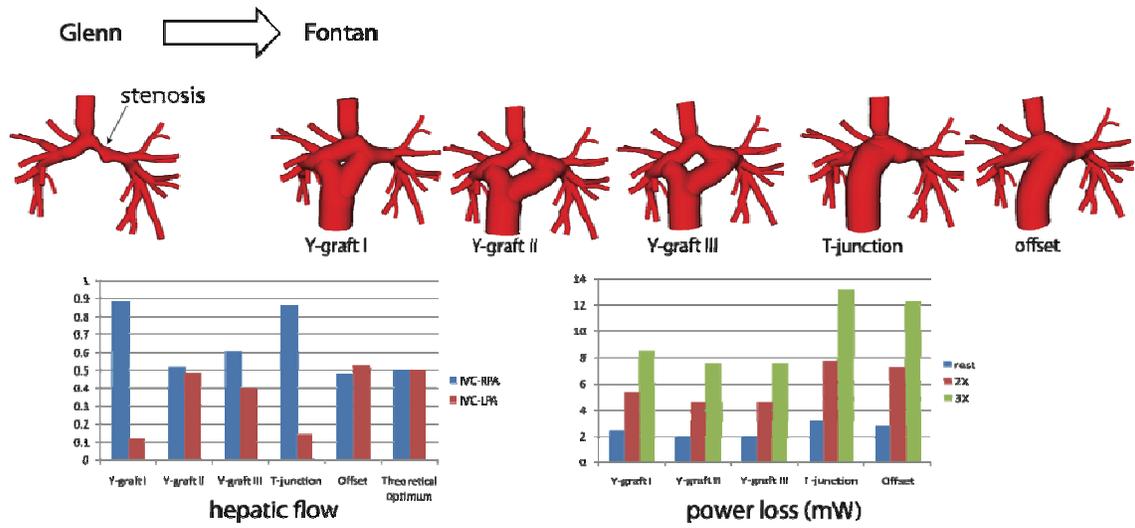

Figure 9: Evaluation of a new Fontan Y-graft design using virtual surgery tools. Performance of the Y-graft, offset, and t-junction designs are compared by examining the hepatic flow distribution and the power loss.

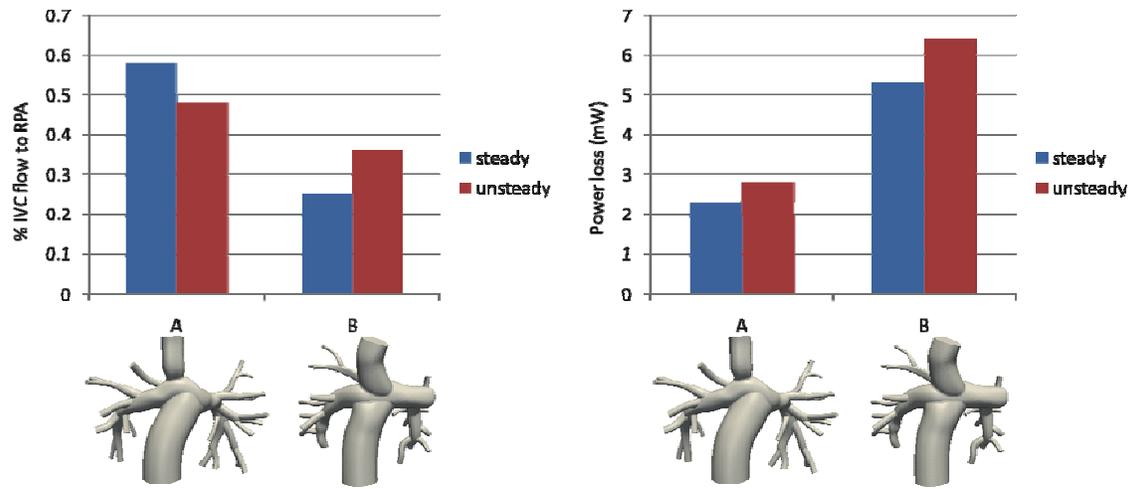

Figure 10: Comparison of predicted hepatic flow distribution using steady inflow vs. pulsatile inflow conditions. For both patients, results differ significantly, demonstrating the importance of boundary conditions in affecting flow results.